\centerline{\bf BRANES FROM MOYAL DEFORMATION QUANTIZATION}
\centerline{\bf OF GENERALIZED YANG MILLS THEORIES}
\bigskip
\centerline{ Carlos Castro}
\centerline{Center for Theoretical Studies of Physical Systems} 
\centerline{ Clark Atlanta University, Atlanta, GA. 30314, USA}
\bigskip
\centerline{August,  1999} 
\bigskip

\centerline { \bf ABSTRACT}
\bigskip
It is shown that a Moyal deformation quantization of the $SO(4k)$ Generalized 
Yang-Mills (GYM) theory action in $D=4k$ dimensions, for spacetime independent 
field configurations, in the $\hbar \rightarrow 0$ limit, yields the 
Dirac-Nambu-Goto $p$-brane actions ( obtained from the conformally invariant Dolan-Tchrakian $p$-brane actions after elimination of the auxiliary world volume metrics) , 
in the orthonormal gauge, for $p+1=4k$ world volumes embedded in a $D=4k$ target spacetime background. The gauge fields/target spacetime coordinates correspondence is required but no large $N$ limit is necessary. 
The equivalence between Moyal SDYM and Self Dual $p$-branes is proposed without
choosing the orthonormal gauge. 
\bigskip

\centerline{\bf I. {INTRODUCTION}}
\bigskip

Recently, conformally invariant sigma models in $D=2n$ dimensions with target
 non-compact $O(2n,1)$ groups were studied. Instanton field configurations were found to correspond geometrically to conformal `` stereographic'' mappings of 
$R^{2n}$ into Euclidean signature $AdS_{2n}$ spaces [1]. These conformally invariant sigma models [2] were of crucial importance in the construction of conformally invariant Lagrangians, with vanishing world volume cosmological constant, for 
bosonic $p$-branes, such that $p+1=2n$ [3]. 
Closely related to these conformally invariant sigma models are the $SO(4k)$ Generalized Yang Mills Theories (GYM) in $R^{4k}$ constructed by Tchrakian [4]. Hierarchies of gauged Grassmanian Models in $4k$ dimensions with Self Dual Instantons were also constructed by Tchrakian and Manvelyan [4].  

Nahm equations are directly connected with Self Dual Yang Mills equations. Moyal Deformation Quantization of Nahm equations,  and the Toda models in the $N\rightarrow \infty$ limit, were constructed in [5] and solutions were found in [6] in relation to Matrix models and $M$ theory. Deformation quantization has recently captured a lot of interest in string theory and for this reason we will construct the Moyal Deformation Quantization of Tchrakian $SO(4k)$ GYM in $R^{4k}$ and show that in the $\hbar \rightarrow 0$ limit, and for spacetime independent field configurations, one will have $p$-brane actions, 
in the orthonormal gauge, for $p$-branes, $p+1=4k$, moving in target spacetime backgrounds of dimensionality equal to $4k$. 

Fairlie [6] has emphasized the importance of the large $N$ limit of $SU(N)$ 
Yang-Mills theory , for spacetime independent field configuartions, in connection with the Schild' string action. Fairlie has raised the interesting issue that  a possible extension of the large $N$ limit to Dirac-Born-Infeld actions, 
$D$-Branes effective actions. may also exist. This would pave the road, in a very natural fashion, to a deformation quantization program to $D$-branes based on Dirac-Born-Infeld actions.   

The standard work on Deformation Quantization was initiated by  [7,8,9]. For the role of Fedosov Deformation Quantization in $W_\infty$ Geometry [10]; on the connection between holography and the quantum geometry of surfaces with Moyal quantization see
[11]. The importance of $W_\infty$ symmetry in the Moyal quantization process
was studied in  [12]. For recent results on $D$-Branes and Deformation Quantization and an extensive list of current references on Star products and string theory see [13].  A rigorous mathematical foundation is given in [14] and a  path integral expression of Kontsevich quantization was presented in [15]. Finally, Deformation Quantization of coadjoint orbits in semisimple groups we refer to [16]. Geometric induced actions for $W_\infty$ gravity, an anomalous 
effective WZNW action, based on a coadjoint orbit method associated with the Moyal deformations of the algebra of 
differential operators of the circle was given by Nissimov, Pacheva and 
Vaysburd [16].

We start with a few definitions of GYM in the next section before performing the Moyal quantization that will allow us to show its relation to the 
construction of $p$-brane actions. Finally we present our conclusions.

\bigskip
\centerline{\bf II. Deformation Quantization of GYM}
\bigskip
The Generalized Yang Mills Theory (GYM) described by Tchrakian [4] based on $SO(4k)$ in $R^D$ where 
$D=4k$ has for defining  Lagrangian :

$$L={1\over g^2} tr~(F^{\alpha_1\alpha_2......\alpha_{2k}}_{\mu_1\mu_2.......\mu_{2k}}
\Sigma_{\alpha_1\alpha_2.......\alpha_{2k}}       )^2 = {1\over g^2} tr~(F_2 \wedge F_2......F_2)^2. \eqno (1)$$
$g$ is a dimensionless coupling constant and   
$$\Sigma_{\alpha_1\alpha_2.......\alpha_{2k}}=\Sigma_{[\alpha_1\alpha_2}  
\Sigma_{\alpha_3\alpha_4}.......\Sigma_{\alpha_{2k-1}\alpha_{2k}]}.\eqno (2)$$
is the totally antisymmetrized product of $k$ factors of the 
$2^{2k-1}\times 2^{2k-1}$ matrices $\Sigma_{\alpha\beta}$ corresponding to the chiral representation of $SO(4k)$.   
For $k=1$ one has the usual $SO(4)\sim SO(3)\otimes SO(3)$ , whose double cover is $SU(2)\otimes SU(2)$ : 

$$\Sigma_{\alpha\beta}=-{1\over 4}\Sigma_{[\alpha} {\tilde \Sigma}_{\beta]}. \eqno (3)$$
$$\Sigma_\alpha ={1\over 2} (1+\gamma_5)\gamma_\alpha.~~~
{\tilde \Sigma}_\beta  ={1\over 2} (1-\gamma_5)\gamma_\beta.~~~\alpha, \beta =1,2,3,4. \eqno (4)$$

For other values of $k$ one has :

$$\Sigma_{\alpha\beta}=-{1\over 4}\Sigma_{[\alpha} {\tilde \Sigma}_{\beta]}.$$
$$\Sigma_\alpha ={1\over 2} (1+\gamma_{4k+1})\gamma_\alpha.~~~
{\tilde \Sigma}_\beta  ={1\over 2} (1-\gamma_{4k+1})\gamma_\beta.~~~\alpha, \beta =1,2.......4k \eqno (5)$$

A Moyal deformation quantization of the $SO(4k)$ GYM assigns a one-to-one map of the self-adjoint tensor-valued operators, 
$\hat F_{\mu_1.....\mu_{2k}}$, belonging to the Hilbert space of square integrable functions, $L^2 (R^{2k})$, to the $c$-number tensor-valued functions of the phase space $R^{4k}$ coordinates. The deformation quantization also assigns a one-to-one map from the commutators of self-adjoint operators  $[{\hat A}, {\hat B}]$ to the Moyal brackets 
$\{\{A, B\}\}$ where $A =symbol [{\hat A}]$ and $B=symbol [{\hat B}]$.  The symbol of a self-adjoint operator ${\hat A} $, which assigns a $c$-number function of phase space, is defined via the Weyl Groenewold Wigner Moyal map (WWGM) [7]  : 

$$A (x^\mu, q^1,p^1,.....q^{2k}, p^{2k}) \equiv \int d^{2k}\xi <{\vec q}-{ {\vec \xi}\over 2} | {\hat A}(x^\mu)|{\vec q}+{ {\vec \xi} \over 2} >
e^{  {i{\vec p}.{\vec \xi}\over \hbar}}. \eqno (6)$$
In order to evaluate the $matrix$ elements of  integrand of (6) one needs to know the representation of the $SO(4k)$ algebra in the Hilbert space $L^2 (R^{2k})$. However,  for our purposes at the moment this will not be necessary. THE WWGM map (6) was used by Fairlie to construct solutions to the Moyal Nahm equations in $3,7$ dimensions. 

The WWGM map associated with ${\hat F_{\mu_1\mu_2......\mu_{2k}}}$ will be 
the $c$-number tensor-valued function of the phase space variables, in addition to the 
$D=4k$ spacetime $x^\mu$ coordinates, :

$$ F_{\mu_1\mu_2......\mu_{2k}} (x^1,x^2,.....x^{4k}; q^1,p^1,.........q^{2k}p^{2k}). \eqno (7)$$
From the definition given by (1,2) it can be written as the antisymmetrized product of the $k$ factors of :

$$F_{\mu_1\mu_2......\mu_{2k}}=F_{[\mu_1 \mu_2} F_{\mu_3 \mu_4}........
F_{\mu_{2k-1}\mu_{2k}]}. \eqno (8)$$
with :

$$F_{\mu\nu}=\partial_\nu  A_\mu - \partial_\mu A_\nu +\{\{A_\mu, A_\nu\}\}. .\eqno (9 ) $$
For the particular case of space-time independent field configurations the above field strengths become :
$$F_{\mu_1\mu_2......\mu_{2k}}= \{\{A_{\mu_1}, A_{\mu_2} \}\}\wedge     
\{\{   A_{\mu_3}, A_{\mu_4} \}\}\wedge .......................
\wedge  \{\{ A_{\mu_{2k-1}}, A_{\mu_{2k}} \}\}. \eqno (10)$$
The Moyal star product  in the $R^{4k}$ phase space is defined as :

$$A*B = \sum^\infty_{k=0} ({i\hbar\over 2})^k {1\over k!} \omega^{i_1j_1}\omega^{i_2 j_2}.....\omega^{i_k j_k} (\partial_{i_1 i_2.....i_k} A) 
(\partial_{j_1 j_2.....j_k} B).\eqno (11a)$$
and the Moyal bracket is 
$$\{\{ A, B\}\} = {1\over i\hbar} (A*B -B*A) \eqno (11b) $$
where $\omega^{ij}$ are the components of the inverse tensor of the symplectic 
two-form $\omega_{ij}$. In the classical $\hbar \rightarrow 0$ limit, the Moyal bracket reduces to the Poisson bracket.  The trace over the Lie algebra indices is converted, after the WWGM prescription,  into an $integral$ over the $flat$ phase space coordinates. The relevance of the Moyal quantization, and a subsequent $\hbar \rightarrow 0$ limit, is that $no$ large $N\rightarrow \infty$ limit is required in the associated $SO(2N),SU(N)$ Yang Mills theories. for example.  No $\infty \times \infty$ matrices are needed. The $\hbar \rightarrow 0$ limit will render the Poisson bracket area-preserving diffs algebra automatically. See Garcia-Compean 
et al  [5].

When the phase space is $curved$ the Moyal deformation quantization needs to be replaced by a deformation quantization program described by  Fedosov [9] that takes into account the fact that the ordinary Moyal star product in curved phase space is no longer associative.  A geometric derivation of $W_\infty$ gravity from Fedosov deformation quantization and the reason why the string coordinates inherit a noncommutative product structure, among other things, was given in [10].

Now, taking into account the special case of independent spacetime field configurations, the action associated with  the Moyal quantization of the $SO(4k)$ GYM in $R^{4k}$ factorizes into a product of a space-time volume and an integral over phase space variables. corresponding to the trace over the Lie algebra elements  : 

$${\Omega_{4k}\over g^2}\int d^{2k}q ~d^{2k} p~ [\{\{A_{\mu_1}, A_{\mu_2} \}\}\wedge     
\{\{   A_{\mu_3}, A_{\mu_4} \}\}\wedge .......................\wedge
 \{\{ A_{\mu_{2k-1}}, A_{\mu_{2k}} \}\}]^2. \eqno (12)$$

The classical limit $\hbar =0$ of the Moyal bracket yields the Poisson bracket represented by a single bracket symbol. Hence, in the $\hbar=0$ limit the action for the space time independent field configurations becomes : 

$${\Omega_{4k}\over g^2}     \int d^{2k}q ~d^{2k} p ~[ \{A_{\mu_1}, A_{\mu_2} \}\wedge     
\{A_{\mu_3}, A_{\mu_4} \}\wedge .......................\wedge\{ A_{\mu_{2k-1}}, A_{\mu_{2k}} \}]^2. \eqno (13)$$
If one equates the flat  phase space coordinates, $q^1,p^1,.....q^{2k}, p^{2k}$ with the coordinates of the $4k$-dimensional world volume of a $p$-brane such that $p+1=4k$ : $\sigma^1=q^1, \sigma^2 =p^1,....\sigma^{4k-1} =q^{2k}, \sigma^{4k}=p^{2k}$  and establishes the spacetime independent gauge fields/ coordinate correspondence : 
$A_\mu (\sigma) \leftrightarrow {1\over l^2_P} X_\mu (\sigma)$  , with $l_P$ the Planck length in the appropriate dimension, 
we will see that the above action (13) is nothing but the  Dirac-Nambu-Goto (DNG) action for a $p$-brane moving in  target spacetime backgrounds of dimension $4k$, for the values of $p+1=4k$ in the orthonormal gauge. Let us take without loss of generality the case $k=1$ associated with a $p=3$-brane whose world volume if four dimensional. The $\hbar \rightarrow 0$ limit yields in (13)  :

$${\Omega_{4} \over g^2 l^8_P } \int d^{4}\sigma ~[ \{X_{\mu_1}, X_{\mu_2} \}]^2 =
{\Omega_{4} \over g^2 l^8_P} \int d^{4}\sigma ~ \omega^{i_1 j_1} 
\omega^{i_2 j_2}\partial_{i_1}  X^{\mu_1} \partial_{j_1 } X^{\mu_2}
 \partial_{i_2}  X_{\mu_1}\partial_{j_2 } X_{\mu_2}. \eqno (14a)$$
where $\omega^{i j}$ is the inverse of the  symplectic form used to define the Poisson bracket. In a canonical basis the action in (14a) becomes   
$$
{\Omega_{4} \over g^2 l^8_P} \int d^{4}\sigma ~~ 
[(\partial_{[q_1}  X^{\mu_1} \partial_{p_1] } X^{\mu_2}) +
 (\partial_{[q_2}  X^{\mu_1}\partial_{p_2] } X^{\mu_2})]^2. \eqno (14b)$$

The conformally invariant Dolan-Tchrakian action [3] for a $p=3$-brane whose world volume is four dimensional embedded in a  four dimensional target spacetime background is :

$$ T \int d^{4}\sigma ~{ \sqrt h}~ h^{\alpha_1 \beta_1} h^{\alpha_2 \beta_2}\partial_{[\alpha_1}  X^{\mu_1} \partial_{\alpha_2 ]} X^{\mu_2}
 \partial_{[\beta_1}  X_{\mu_1}\partial_{\beta_2 ]} X_{\mu_2}. 
\eqno (15)$$
The action is conformal invariant under the transformations of the world volume auxiliary metric : 

$$h_{\alpha_1 \alpha_2} \rightarrow e^{\Omega} h_{\alpha_1 \alpha_2}. ~~~
h^{\alpha_1 \alpha_2} \rightarrow e^{-\Omega} h^{\alpha_1 \alpha_2}. ~~~
{\sqrt h} \rightarrow e^{2\Omega} {\sqrt h}. \eqno (16)$$

As previously said, the Moyal quantization requires a flat phase space which seems to be incompatible with the action (15) for arbitrary world volume metrics, $h_{\alpha \beta}$. This can be solved in a two step process. Firstly, by noticing that the action (15) is on-shell equivalent to the Dirac-Nambu-Goto actions (DNG)  upon the elimination of the metric $h_{\alpha \beta}$ and plugging its value back into the action (15) to retrieve the action :

$$S_{DNG}=T\int d^4\sigma~ \sqrt { det~G_{\alpha\beta } } .~~~ G_{\alpha\beta}=
\partial_\alpha X^\mu \partial_\beta X_\mu. \eqno (17)$$
where $G_{\alpha\beta}$ is the induced world volume metric as a result og the embedding of the $p=3$-brane into the target spacetime background. To verify this result the authors [3] have shown that the algebraic elimination of $h_{\alpha \beta}$ yields one physical solution given by :  

$$ w^{-1} G_{\alpha\beta}=h_{\alpha\beta}.~{\sqrt h} =w^{-2}{\sqrt G}. 
\eqno (18)$$
and , as expected,  the action (15) density  for the solutions of $h_{\alpha \beta}$ becomes proportional to the ${\sqrt  G }$. The second solution [3] was disregarded on physical grounds. 
We have seen then how the on-shell world volume metric $h_{\alpha \beta}$ is 
in the same conformal class to the embedding metric $G_{\alpha\beta}$.

In the second step one fixes the $4$-dimensional ( Euclidean) world volume reparametrization invariance of the DNG action ( obtained after the elimination of the auxiliary metric) by choosing the analog of the orthonormal gauge in the Dirac-Nambu-Goto string ( for Euclidean world sheets) : 

$$ (\partial_\alpha X^\mu)(\partial_\beta X_\mu) =\delta_{\alpha \beta}.
~~~\alpha, \beta =1,2,3,4.  \eqno (19)$$
one could have included a scalar factor in front of $\delta_{\alpha \beta}$ 
in (19) if one wishes without changing the final results below. Hence, one arrives at the conclusion that the on-shell solutions  to the auxiliary world volume metric, in the orthonormal gauge, 
are given by conformally flat (Euclidean) metrics : $h_{\alpha_1 \alpha_2} =  e^{\Omega} \eta_{\alpha_1 \alpha_2}$ where $\eta_{\alpha_1 \alpha_2} $ is the  (Euclidean) four dimensional world volume metric. The above action (15) 
becomes finally  :

$$ T \int d^{4}\sigma ~ \eta^{\alpha_1 \beta_1} \eta^{\alpha_2 \beta_2}\partial_{[\alpha_1}  X^{\mu_1} \partial_{\alpha_2 ]} X^{\mu_2}
 \partial_{[\beta_1}  X_{\mu_1}\partial_{\beta_2 ]} X_{\mu_2}=  
T \int d^{4}\sigma ~ 
[\partial_{[\alpha_1}  X^{\mu_1} \partial_{\alpha_2 ]} X^{\mu_2}]^2 
\sim T\int d^4\sigma \eqno (20) $$
the Lagrangian density of (20) becomes the sum of squares, which in the orthonormal gauge (19) turns into a $constant$, as it should,  since in the orthonormal gauge (19) the measure density in the DNG action (17) becomes equal to $1$. This is indeed compatible with choosing a flat phase space
and in equating it with the world volume of the $p=3$-brane.

Similar arguments follow with eqs-(14a, 14b), after making the identification : $\sigma^1=q^1, \sigma^2 = p^1, \sigma^3 =q^2, \sigma^4 =p^2$, the Lagrangian density in (14b) also becomes a sum of squares, since in the orthonormal gauge (19) the cross terms cancel, and it yields finally a $constant$ like it occurred in (20). 
 The fact that one can gauge the Lagrangian density to a $constant$ is related  to the fact that a $p$-brane moving in a background of $D=p+1=4k$ dimensions does not have local transverse degrees of freedom : 
$D-p-1=0$. In that sense the theory is topological.  Before continuing, we must emphasize that the action (14) is $not$ the Schild action because the latter is the action for a string, a two-dim world sheet, whereas the action in (14) is truly a four-dim one. The Schild action is area-preserving diffeomorphisms invariant but it is not fully reparametrization invariant like the Dolan-Tchrakian $p$-brane actions.

When the target spacetime dimension is saturated, $D=p+1=4k$, the DNG action becomes trivial :

$$S_{DNG} =T\int d^4\sigma~{\partial (X^1,X^2,X^3,X^4)\over \partial 
(\sigma^1, \sigma^2, \sigma^3, \sigma^4)}. \eqno (21)$$
where the measure of integration is simply the Jacobian of the change of variables from 
$X^\mu $ to $\sigma^i$. A trivial solution to the 
orthonormal gauge conditions (19) is given by :

$$X^1 =\sigma^1.~~X^2 =\sigma^2.~~X^3 =\sigma^3.~~X^4 =\sigma^4. \eqno (22)$$
which renders the Jacobian equal to unity and this is compatible with the 
flatness of the phase space. A curved phase space description requires a Fedosov deformation quantization as stated earlier.

Therefore, to sum up, the  action (14) obtained from a Moyal deformation quantization of a $SO(4)$ Yang Mills Theory in $R^4$, in the $\hbar \rightarrow 0$ limit,  for spacetime independent field configurations, 
is equivalent to DNG actions ( obtained from the Dolan-Tchrakian action (15) after the elimination of the auxiliary world volume metric $h_{\alpha \beta}$)  , in the orthonormal gauge,  if, and only if, one equates 
the $p=3$-brane tension ( up to a numerical factor)  to :
$T_3\sim  {\Omega_4 \over g^2 l^8_P}.$
$\Omega_4$ is the four dimensional spacetime volume of $R^4$ where 
the $SO(4)$ Yang Mills theory is defined. To regularize quantities like the volume  we could select a compact spacetime. Similar conclusions follow had one introduced a scalar factor in front of the $\delta_{\alpha \beta}$ when the orthonormal gauge was chosen.    

It is important to emphasize that $X_\mu\not= x_\mu$. The base manifold coordinates over which one defines the $SO(4k)$ GYM must $not$ be confused with the target spacetime embedding coordinates of the $p$-brane ($p+1=4k$). 
The gauge fields $A_\mu$, after the Moyal quantization and the $\hbar =0$ limit, are the ones that behave like the $p$-brane coordinates. In general, the $p$-brane tension, where $p+1=4k$, up to a $k$-dependent numerical constant,  is then :
$T_p \sim  { \Omega_{4k} \over g^2 l^{8k}_P}. $
and the Moyal deformation quantization of the $SO(4k)$ Generalized Yang Mills theory (GYM), for spacetime independent  field configurations, in the $\hbar \rightarrow 0$ limit, is :  

$${\Omega_{4k} \over g^2 l^{8k}_P } \int d^{4k}\sigma ~[ \{X_{\mu_1}, X_{\mu_2} \}\wedge     
\{X_{\mu_3}, X_{\mu_4} \}\wedge .......................\wedge  \{ X_{\mu_{2k-1}}, X_{\mu_{2k}} \}]^2. \eqno (23)$$
Similar arguments as the $k=1$ case will allow us to show that the action given by (23) will be equivalent to the  DNG actions ( after the elimination of the world volume metrics from the Dolan Tchrakian $p$-brane action) , with $p+1=4k$,  moving in  target spacetime backgrounds of $D=4k$, 
in  the orthonormal gauge (19),  where the world volume and spacetime indices now run over $4k$ values  :

$$ T \int d^{4k}\sigma ~{ \sqrt h}~ h^{\alpha_1 \beta_1} h^{\alpha_2 \beta_2}
........h^{\alpha_{2k}\beta_{2k}} 
\partial_{[\alpha_1}  X^{\mu_1} \partial_{\alpha_2 } X^{\mu_2}.....
\partial_{\alpha_{2k} ]}X^{\mu_{2k}}
\partial_{[\beta_1}  X_{\mu_1} \partial_{\beta_2 } X_{\mu_2}.....
\partial_{\beta_{2k} ]}X_{\mu_{2k}}. \eqno (24)$$

Another less $restrictive$ alternative to the one presented above involving the elimination of the auxiliary metric and choosing the orthonormal gauge is to write the action (24) in geometric terms [3] which has precisely the same form as a GYM theory :

$$\int d^{4k}x~ F_{\mu_1......\mu_{2k} } \wedge ^* F_{\mu_1.....\mu_{2k}}  
\leftrightarrow T\int d^{4k}\sigma ~(E^{i_1}\wedge E^{i_2}\wedge.....\wedge 
E^{i_{2k}})\wedge ^*
(E^{j_1}\wedge E^{j_2}\wedge....\wedge E^{j_{2k}})\eta_{i_1j_1}\eta_{i_2 j_2}.....\eta_{i_{2k} j_{2k}}.$$
$$tr \leftrightarrow \int dq^{2k}dp^{2k}     \eqno (25)$$
where one has written the Dolan-Tchrakian action purely in geometric terms given by the pullback of the orthonormal one-forms, $E^i$,  for the background $4k$ spacetime onto the $4k$-dim world volume . The Hodge dual star operation is given in terms of the world volume metric $h_{\alpha \beta}$. It is precisely in the star
operation where the metric is encoded. The Hodge star dual operation in the GYM action is defined w.r.t the flat $R^{4k}$ metric as usual.

Since every symplectic manifold can be equipped with an almost complex structure, $J$, the metric tensor defined by : $g(X,Y) = -\omega (JX,Y)$ for all vector fields, $X,Y$, is a Riemannian metric on the symplectic manifold. One can then use the gauge fields/coordinates and symplectic form/metric correspondence :

$$A_\mu dx^\mu \leftrightarrow E_\mu dx^\mu . ~~~\omega^{ij} \leftrightarrow h^{\alpha\beta}. \eqno (26)$$
and ask now : Under what conditions the action of the l.h.s of (25) for arbitrary $\omega^{ij}$,  is equivalent to the r.h.s of (25) after using the one-to-one gauge field/coordinates correspondence (26); i.e. under what conditions, the Moyal deformation quantization of a $SO(4)$ GYM in $R^4$, in the $\hbar =0$ limit, for spacetime independent field configurations, yields the Dolan-Tchrakian actions $without$ having to fix the orthonormal gauge and $without$ having to eliminate the metric $h_{\alpha \beta}$ via its algebraic equations of motion ?  

Let us look at the special case of Moyal deformations of 
the Self Dual Yang Mills sector and the corresponding Self Dual $p$-branes. 
For a typical YM ( Generalized YM) theory  
$F\wedge *F$, the Self Dual solutions 
$^*F =F$ solve automatically the YM (GYM) equations of motion due to the Bianchi identities and yield topological invariants $\int F\wedge F$; i.e the winding number. Instanton numbers associated with noncommutative YM theories based on noncommutative topology have been provided by Zois [17]. 

On the r.h.s of the equation (25) the action for a Self Dual $p$-brane, when the target embedding spacetime dimension is saturated, $D=p+1=4k$ turns out to be essentially topological in the sense that it does not posess local degrees of freedom. When $E\wedge E....\wedge E =^*(E\wedge E....\wedge E)$ one can see in (25) that the Dolan-Tchrakian action for a Self Dual $p$-brane becomes the integral of the Jacobian of the change of variables : $X\rightarrow \sigma$ giving finally an integrated volume. Notice once again, that the metric $h_{\alpha\beta}$ is naturally encoded in the Hodge dual star operation. There is no need to eliminate it.  
The volume form $\Omega$ can be written as a series of wedge products of symplectic two-forms : $\Omega =\omega \wedge \omega.....\omega$. If the area is preserved then the volume is also. The converse is $not$ true, a volume may be preserved but the not necessarily the individual area-components.

How can a volume be a topological invariant ? One can have an infinite family of closed $p$-branes with fixed topology but arbitrary volumes. The answer lies in the fact that it is well known (to the experts) that $p$-branes are 
essentially gauge theories of $volume$ preserving diffs [20] . For example, $su(\infty), w_\infty, 
w_{1+\infty}$ are the classical area-preserving diffs algebras of a sphere, plane, cylinder respectively.  The topology of the two-dim surfaces is clearly sensitive to the type of infinite-dim Lie algebras involved. Hence  it is in this context of gauge theories of $volume$ preserving diffs that the volume can be seen as a conserved ``charge'' in the same vein that the conservation of electric charge in Electromagnetism is a consequence of a local $U(1)$ gauge invariance. In the l.h.s of (25) we have a global conserved topological charge ( for the special case of self dual GYM theories ) whereas in the r.h.s of (25), for Self Dual $p$-branes,  we have a local conserved charge due to the 
volume-preserving diffs gauge invariance. Rigorously speaking, one should speak of a $duality$  between the Self Dual GYM theories and Self Dual $p$-brane actions in eq-(25). This is more reminiscent of Maldacena's $AdS/CFT$ duality.

It is important to study first the proposed equivalence between Moyal deformations of SDYM ( SD Generalized YM), in the $\hbar =0$ limit, for spacetime independent field configurations, to the Self Dual $p$-branes. This occurs $without$ imposing any gauge conditions whatsoever nor eliminating the world volume metrics . This will be a first test of the validity of the gauge fields/coordinates correspondence provided by eq-(25). In this integrable sector of self dual $p$-branes and self dual GYM (instantons) the equivalence provided by the Moyal deformation quantization maybe exact. The second test will be in extending this result beyond the self dual integrable sector.

\bigskip

\centerline{\bf CONCLUSIONS}
\smallskip
After showing the equivalence of the Moyal deformation quantization of the GYM and $p$-branes, under certain special conditions, spacetime independent field configurations, $\hbar =0$ limit....it is worth exploring other cases besides the $D=4k$. In particular, the Zariski deformation quantization of Nambu-Poisson 
Mechanics associated with general values of $p$ ( for all $p-$branes ) that has been studied by Dito et al [18] in connection with gauge theories of volume preserving diffs warrants a further investigation.  The Dirac-Nambu-Goto action for all $p$-branes admits an straightforward deformation by simply performing the Zariski deformation of the Nambu bracket [18] which appears in the definition of the action : square root of the determinant of the induced metric and the latter is just 
the sum of the squares of Jacobians (Nambu brackets). 
An starting point to the quantization of $p$-branes based on $p$-brane Quantum Mechanical Wave equations associated with gauge theories of volume preserving diffs has been given in [19].

The $p$-brane tensions appearing in this work have a typical stringy instanton behaviour ${1\over g^2}$. It is warranted to study how $D$-branes fit in this formalism of Moyal-Fedosov Deformation Quantization. The $D$-brane tensions have a ${1\over g}$ behaviour. The WZNW actions associated with open-strings moving in group manifolds  
with an antisymmetric rank two tensor background, and Dirac-Born-Infeld effective actions  discussed in [6,13] should be a starting point.

Finally, more general cases ought to be investigated other than the restricted case studied here of spacetime independent field configurations; the fact that the number of spacetime dimensions $4k$ equals that of the phase space variables suggests a $complexification$ of the $4k$-world volume coordinates and background spacetime and, correspondingly,  of the gauge fields. This deserves to be studied in the future, especially in relation to the role that symplectic and complex geometry plays in the theory of branes. Fedosov deformation quantization and the generalizations of $W$ Geometry [10] will then be an essential ingredient for more general curved phase spaces.     

\bigskip
\centerline{\bf Acknowledgements}
\smallskip

We wish to thank Valeria Ferrari, Carmen Nu${\tilde n}$ez for their help and hospitality at the University of Buenos Aires, Argentina where this work was completed.  
\bigskip
\centerline{\bf REFERENCES}
\bigskip

1. C. Castro : `` Conformally Invariant $\sigma$ Models on $AdS$ spaces, Chern-
Simons $p$-branes 

and $W$ Geometry `` hep-th/9906176. to appear in Nuc. Phys. {\bf B}.

2. B. Dolan, D.H.  Tchrakian : Phys. Letts. {\bf B 198} (4) (1987) 447. 
 
3. B. Dolan, D.H. Tchrakian : Phys. Letts. {\bf B 202} (2) (1988) 211.

4. D. H. Tchrakian : Jour. Math. Phys. {\bf 21} (1980) 166. 

5. C. Castro : Phys. Letts {\bf B 413} (1997) 53. 

C. Castro, J. Plebanski : Jour. Math. Phys. {\bf 40} (8) (1999) 3738. 

H. Garcia-Compean, J. Plebanski, M. Przanowski : `` Geometry associated with 

SDYM and the chiral approaches to SDG `` hep-th/9702046. 

H. Garcia-Compean, J. Plebanski : `` The Weyl Wigner Moyal Description of $SU(\infty)$ 

Nahm Equations `` hep-th/9612221. 

6. D. Fairlie : `` Moyal Brackets in $M$ Theory `` hep-th/9707190. Mod. Phys.

{\bf A 13} (1998) 263. `` Dirac-Born-Infeld Equations `` hep-th/9902204. 

L. Baker, D. Fairlie : `` Moyal Nahm Equations `` hep-th/9901072 

7. J. Moyal : Procc. Cam. Phil. Soc. {\bf 45} (1945) 99.
E. Wigner : Phys. Rev. {\bf 40} (1932) 749. 

H. Groenewold : Physica {\bf 12} (1946) 405. 
H. Weyl : Z. Physik {\bf 46} (1927) 1. 

8. F. Bayen, M. Flato, C. Fronsdal, A. Lichnerowicz, D. Sternheimer : 

`` Ann. Phys. {\bf 111} (1978) 61. 

9. B. Fedosov : Jour. Diff. Geometry {\bf 40} (1994) 213. 

10. C. Castro : `` $W$ Geometry from Fedosov Deformation Quantization'' to appear

in the Jour. of Geometry and Physics, 1999. 

11. A. Granik, G. Chapline : Moyal Quantization, Holography and the Quantum Geometry

of Surfaces `` Special issue of J. Chaos, Solitons and Fractals {\bf 10} (2-3) (1999). 

12. T. Deroli, A. Verciu : J. Math. Phys {\bf 38} (11) (1997) 5515. 

E. Gozzi, M. Reuter : Int. Jour. Mod. Phys {\bf A 9} (32) (1994) 5801.  

13. V. Schomerus : `` D-Branes and Deformation Quantization ``. JHEP 06 (1999). 

hep-th/9903205. 

A. Yu Alekseev, V. Schomerus : `` D-Branes in the WZW model `` hep-th/9812193.

A. Yu Alekseev, A. Recknagel, V. Schomerus : `` Noncommutative World Volume Geometries :

Branes on a $SU(2)$ and Fuzzy Sphere''. hep-th/9908040. 

H. Garcia-Compean, J. Plebanski : `` D-Branes on Group Manifolds and Deformation 

Quantization `` hep-th/9907183. 

L. Cornalba, R. Schiappa : `` Matrix Theory Star Products from the Born Infeld 

Action''  hep-th/9907211.

N. Seiberg, E. Witten : talks at the Potsdam 99 String Conference. 

14. M. Kontsevich : `` Operads and Motives in Deformation Quantization `` 
math. QA/9904055. 

``Deformation Quantization of Poisson Manifolds `` 
q-alg/9709040.  

D. Tamarkin : `` Another proof of the Kontsevich Formality Theorem `` 
math. QA/9803023. 

15. A.S Cattaneo, G. Felder : `` A Path Integral Approach to the Kontsevich

Quantization `` math.QA/9902090. 

16. A. Astashkevisch : `` On Fedosov Quantization of Semisimple Coadjoint Orbits `` 

MIT Ph. D Thesis. Math. Dept. 1996.

E. Nissimov, S. Pacheva , I. Vaysburd : `` $W_\infty$ Gravity : a geometric approach. hep-th/9207048. 

17. I. Zois : `` A new invariant for sigma models''. hep-th/9904001.

18. G. Dito, M. Flato, D. Sternheimer , L. Takhtajan : `` Deformation Quantization and Nambu 

Poisson Mechanics''  : hep-th/9602016. 

19. C. Castro : `` $p$-Branes Quantum Mechanical Wave Equations `` 
hep-th/9812189   

20. E. Bergshoeff , E. Sezgin, Y. Tanni and P.K Townsend : Ann. Phys. {\bf 
199} (1990) 340.

\bye